\documentclass[prl,aps,floatfix,nofootinbib,twocolumn,letterpaper,reprint]{revtex4}
\usepackage{graphicx}
\usepackage{amssymb} 
\usepackage{amsmath} 
\usepackage{amsfonts} 
\usepackage{amsbsy}
\usepackage{color}

\def\be{\begin{equation}} 
\def\ee{\end{equation}}

\begin{document} 

\title{Reorientation in newly formed fission fragments}

\author{G.F.~Bertsch}

\affiliation{Department of Physics and Institute of Nuclear Theory,
University of Washington, Seattle, Washington 98915, USA
}
 
\email{bertsch@uw.edu}

\begin{abstract}
The strong quadrupolar component of the Coulomb field between newly formed 
fission fragments can 
affect the internal energy of the fragments and their angular 
momentum.  Previous estimates of these effects
gave contradictory conclusions about their magnitude.  Here we calculate
them by solving the time-dependent Hamiltonian equation for the 
daughter $^{100}$Zr produced in the fission reaction
n + $^{235}$U $\rightarrow$ $^{136}$Te + $^{100}$Zr.  The Hamiltonian was
constructed from the projected angular momentum eigenstates of an aligned 
deformed mean-field configuration.
For typical initial conditions, 
the average angular momentum of the lighter fragment increases by 1.5 - 3 units.
\end{abstract}

\maketitle

{\it Introduction.}
The strong Coulomb field in early post-scission fragment
interactions  could  induce transitions that change
the $J$-population before the neutron and gamma emission
cascade starts.  This question was addressed in early
publications \cite{ho64,wi72} and follow-up studies (eg., \cite{ra69,mi99}) using similar
methods to Ref. \cite{ho64}.  There the orientation of the two fragments
at the scission point was treated by classical mechanics;
the change in angular momentum was attributed to the torque from the 
Coulomb field of the partner fragment.
In Ref. \cite{wi72} the process was treated as 
in heavy-ion reaction theory where the nuclei are excited by
their mutual Coulomb fields \cite{al56}.  These two studies came to opposite
conclusions: Ref. \cite{ho64} found that the generated angular momentum
was comparable to what was observed, but Ref. \cite{wi72} found
that the post-scission Coulomb field had little effect on the
final states of the fragments.  None of the studies to date have
used the improved theoretical tools that we now possess based on 
self-consistent mean-field 
theory (SCMF) \cite{be03}.  In this note we will examine the
effect of the Coulomb field from the partner nucleus on a typical
SCMF configuration of a scission product.  These 
configurations are deformed with the deformation axis aligned along
the fission direction.
As a result they consist of a coherent superposition of 
angular momentum states, each having a vanishing
angular momentum about the fission axis.  We construct Hamiltonian
matrix in that space of angular momentum eigenstates.
The driving term that changes the composition of the
wave function is the time-dependent Coulomb field from the other
daughter nucleus. 

{\it Hamiltonian.}
The Hamiltonian basis is composed of angular momentum eigenstates
defined by projection from the deformed initial configuration.
The  decomposition was carried
out in Ref. \cite{be18} for the purpose of determining the
importance of deformation in the final state and the effect
of the alignment on the gamma decays in the final state.
The probability $P_J$ of  angular
momentum $J$ was found to be very well approximated by the
spin-cutoff model\footnote{See also Ref. \cite{bo07}.}
\be
\label{PJ}
P_J \sim (2 J+1)\exp\left(-J (J +1)/ 2\sigma^2\right).
\ee
We assume that the intrinsic deformed configuration is invariant under 
time reversal and contains only even angular momenta.
The average angular momentum was estimated 
in  Ref. \cite{be18} as
\be
\langle J^2\rangle \approx 0.3 A^{3/2} \beta
\ee
where $A$ in the number of nucleons in the fragment and
$\beta$ is the deformation parameter, which we define in 
terms of the mass quadrupole moment $Q_0 = \langle 2z^2 - (x^2+y^2)\rangle$
as
\be
\label{Q2b}
\beta = \frac{\sqrt{ 5 \pi} A^{5/3}}{3 r_0^2} Q_0 
\ee
with $r_0 = 1.2$ fm. 

The Hamiltonian consists of a rotational term $H^{\rm rot}$ together
with term coupling to the quadrupole field of the partner nucleus
$H^Q$,
\be
H = H^{\rm rot} + H^{Q}.
\ee
The first term has the matrix elements
\be
H^{\rm rot}_{JJ'} =   \left(\frac{\hbar^2 }{ 2 {\cal
I}}\right)J(J+1)\delta_{J,J'}. 
\ee
where the moment of inertia ${\cal I}$ may be estimated\footnote{There
will be an additional small contribution from the orbital angular momentum
of the two fragments about the center of mass.}  
as 
\be
{\cal I} =  \frac{1}{2} {\cal I}_{\rm rig}
\ee
with \cite[Eq. 4-104]{BM}  
\be
{\cal I}_{\rm rig}= \frac{2}{5} A^{5/3} M r_0^2 ( 1 +
\beta/3).
\ee
The coupling to the quadrupole field  at the
daughter nucleus  (and arising from the partner 
nucleus) is mediated by the electric 
quadrupole operator \be
\hat Q_e = e \sum_p  \left(2 z^2_p -x^2_p  - y^2_p\right).
\ee
Its matrix elements are estimated in the rotor model as \cite[Eq. (4-68a)]{BM}
\be
\langle J' | \hat Q_e | J \rangle = 
\frac{1}{10}\sqrt{(2 J + 1)(2 J' +1)} ( J 0 J' 0 | 2 0)^2 Q_e. 
\ee
Here $Q_e$ is the expectation value of the operator in the intrinsic
state.  The approximate relation to the mass quadrupole moment
is
\be
\label{Qe}
Q_e \approx \frac{e Z}{A} Q_0.
\ee

The Hamiltonian matrix elements of the quadrupole interaction are given by
\be
H_{J,J'}^Q = V_Q \langle J' | \hat Q_e | J \rangle
\ee 
where the Coulomb field strength $V_Q$ depends on the separation $R$ between the
centers of mass of the two fragments 
as
\be
\label{Vr}
V_Q(t) = \frac{Z_2 e}{2 R^3}.
\ee
Here $Z_2$ is the charge of the partner nucleus.

The time evolution of $R(t)$ is treated classically
by the Newtonian force equation,
\be
 \frac{d^2 R}{d t^2} =  \frac{M_1+M_2}{ M_1 M_2} \frac{Z_1 Z_2 e^2 }{ R}
\ee
Finally, we solve the time-dependent Hamiltonian equation
\be
\label{Ht}
i \hbar \frac{d \psi}{ d t} = H \psi
\ee
where $\psi$ is the  vector of $J$-amplitudes and $H$ has
the explicit time dependence coming from $V_Q$.     

Before carrying out the numerical solution of the equations of motion,
we note some of the relevant energy scales in $H$. In the main example below,
we will treat a case with a rather large deformation, $\beta=0.5$.  
The initial energy in the quadrupole field is 
\be
E_Q \approx \frac{e Z_1 Z_2 }{ R^3(0)} Q_e \approx 3.5 \,\,\,{\rm MeV}
\ee
to be compared with the rotational energy of the initial configuration,
\be
\sum_J P_J  \left(\frac{\hbar^2 }{ 2 {\cal I}}\right)J(J+1)\approx 4
\,\,\,{\rm MeV}.
\ee
Since the two terms in $H$ are comparable, the dynamical equations should
be integrated numerically rather than relying on a perturbative 
treatment.  However, we should not be surprised if  
the average angular momentum is slow to change.
This follows from the structure of the Hamiltonian
in the limit where $V_Q$ is constant and $H^{\rm rot}$ can be ignored.
Then the Hamiltonian matrix is tridiagonal with matrix elements
approaching   $H_{J,J} \approx V_Q Q_e/4$ and $ H_{J,J+2} \approx 3 V_Q Q_e/ 8$ 
as $J$ becomes large.  This is similar to the discretized Hamiltonian
for a particle in a one-dimensional box.  If the initial 
wave function is smooth and spread over many states, the subsequent evolution 
will be a diffusive expansion of the wave packet rather a displacement
one way or the other.  

In the absence of an external field, the
quadrupole moment will undergo oscillations due to the
changing relative phases of the $J$-states, but without any corresponding
change in their probabilities.  The period of the oscillation can be
roughly estimated as 
\be
\tau = \frac{\pi\cal I}{\hbar\langle J\rangle} \approx 1200 \,\,\,{\rm fm/c},
\ee
which longer than the interaction time scale for $V_Q$.  

\begin{figure}[htb] 
\begin{center} 
\includegraphics[width=\columnwidth]{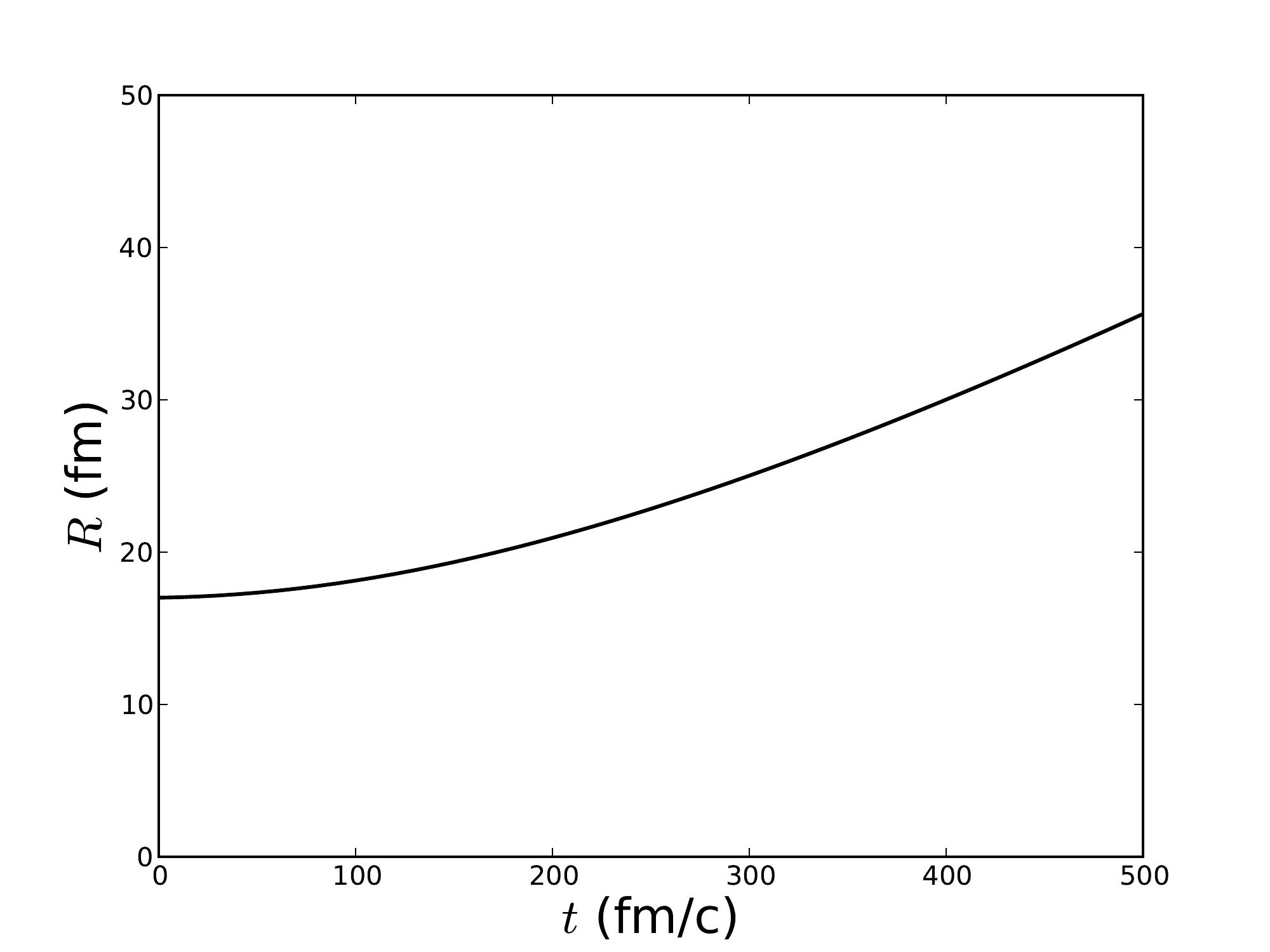}
\caption{ Separation of the two fission fragments as a function of 
time.
\label{coul_acc}
}
\end{center} 
\end{figure}

{\it Application.}
To set the parameter values, we consider a typical fission decay
\be
^{236}{\rm U}_{92} \rightarrow ^{136}{\rm Te}_{52} + ^{100}{\rm Zr}_{40}.
\ee
and examine the evolution of the
$^{100}$Zr$_{40}$ daughter nucleus.
The initial conditions for the separation coordinate are taken as
$R = 17$ fm and $d R/d t =0$ at $t=0$. 
Fig. \ref{coul_acc}  shows the separation of the fragments as a function of 
time. One sees that $V_Q$ will become small on a time scale of a few hundred
fm/c due to its cubic dependence on R$^{-1}$.

The deformation of the
lighter fragment might be very large.  A recent study with time-dependent
mean-field theory \cite{bu18} found deformations in the range
$\beta = 0.55 - 0.7$.  In a pioneering early study of the statistical model of
scission \cite{wi76}, it appeared that $\beta\approx 0.6$ was favored for
the light fragment.  For the present modeling, we
will show details of the Hamiltonian evolution
taking $\beta = 0.5$ in Eq. (2) and (3).  From Eq. (2), the average 
angular momentum is given by $\langle J^2\rangle^{1/2} \approx 12 $.  This is
much larger than estimates based on characteristics
of the neutron- and gamma-decay cascades of the daughter nuclei, so
we will  consider initial states with lower $\langle J \rangle$ as well.  
For the present example, the $J$-amplitudes are truncated
beyond $J=24$, resulting in a 13-dimensional Hamiltonian.
The relative amplitudes in the initial
wave function are taken as 
$a_J = P_J^{1/2}$ from probabilities determined by Eq. (1). For the
other parameters in the Hamiltonian,  
the electric quadrupole moment is $Q_e = 4.7$ e-b 
and the moment of inertia is
${\cal I} = 17.5 $
MeV$^{-1}$ $\hbar^{-2}$ from Eq. (5).
\begin{figure}[htb] 
\begin{center} 
\includegraphics[width=\columnwidth]{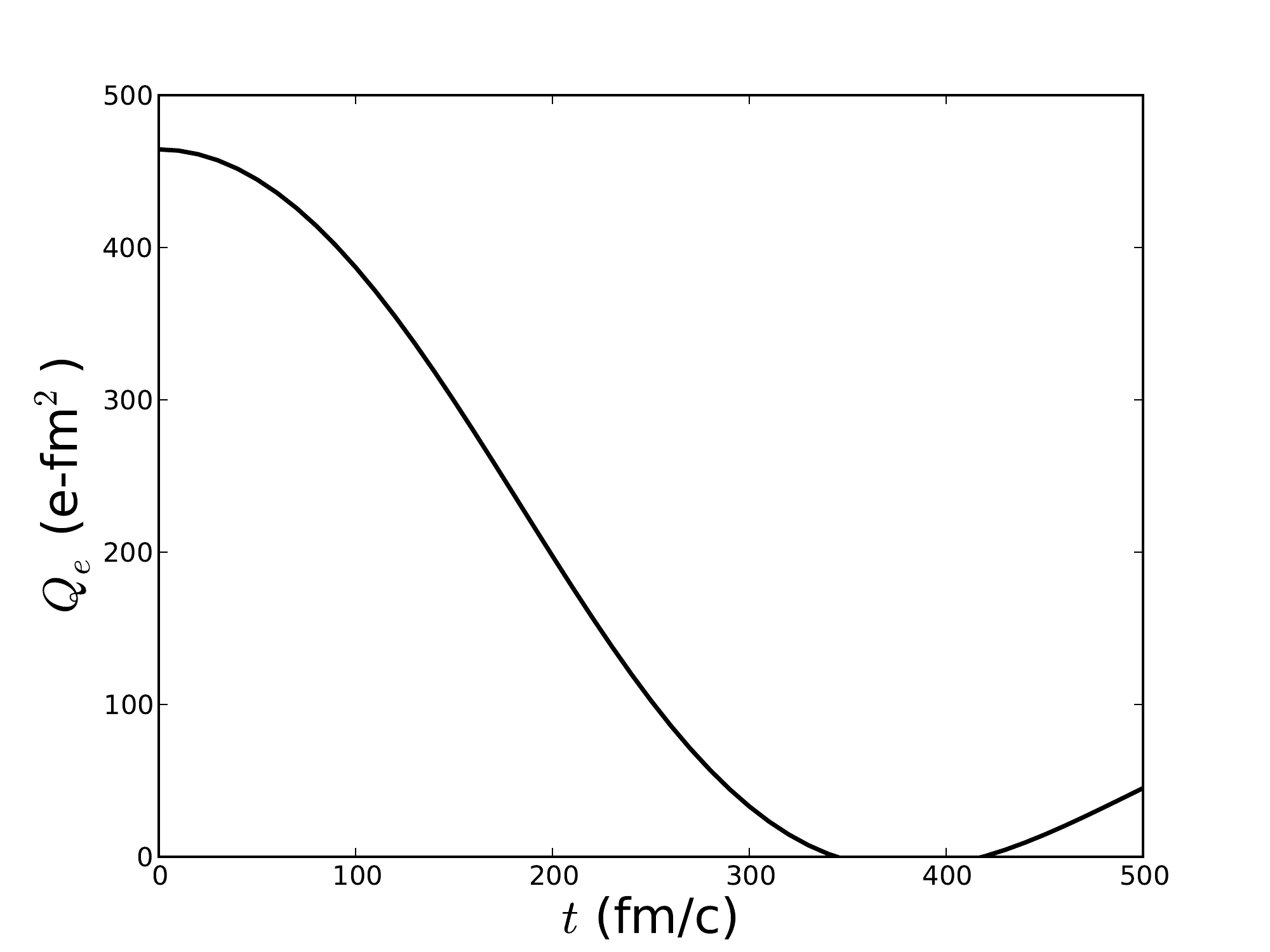}
\caption{ 
\label{Qvst} Electric quadrupole moment of the lighter fragment 
as a function of
time after scission.
}
\end{center} 
\end{figure} 
\begin{figure}[htb] 
\begin{center} 
\includegraphics[width=\columnwidth]{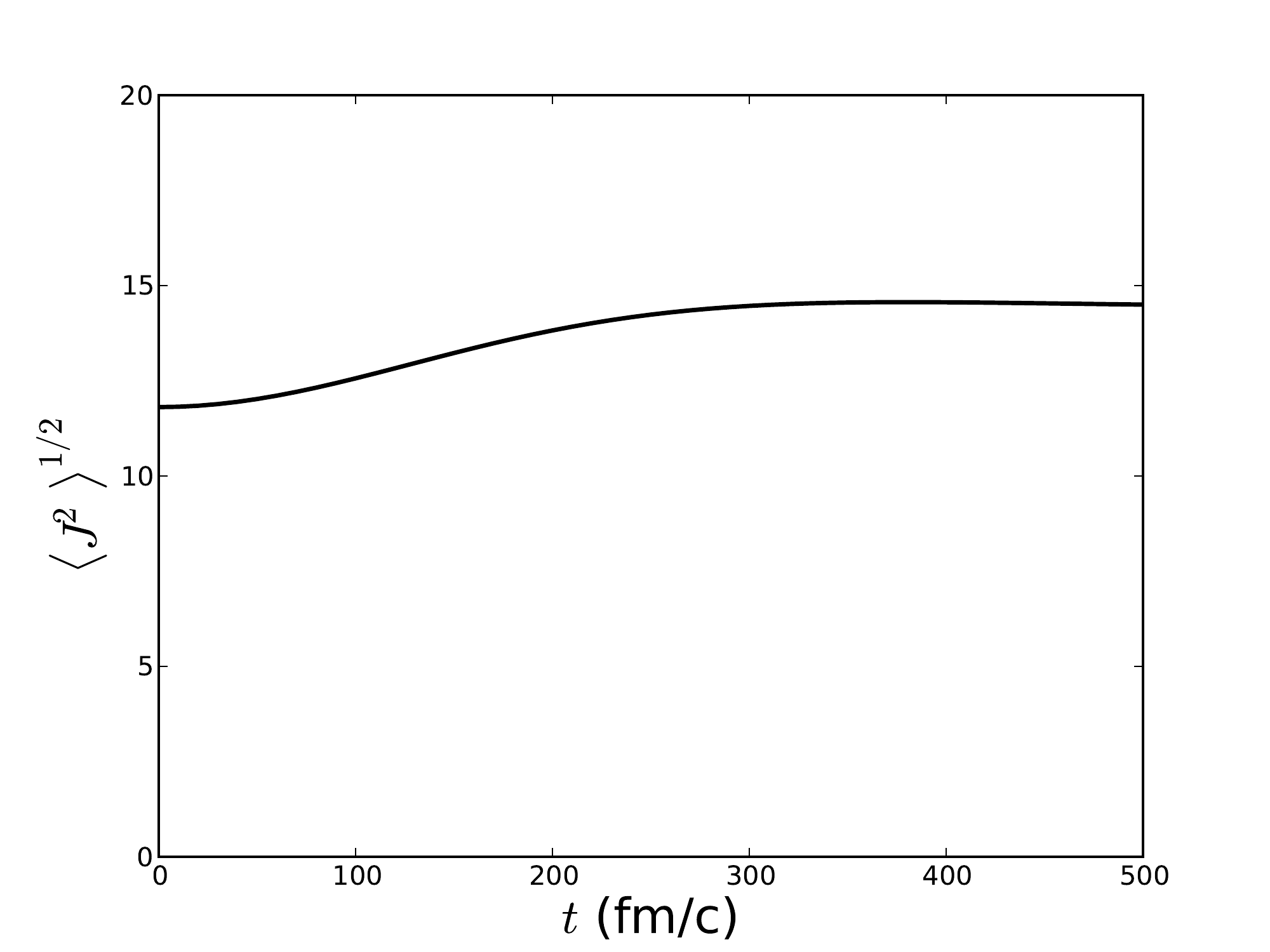}
\caption{ 
\label{J2vst} Mean square angular momentum as a function of
time after scission.
}
\end{center} 
\end{figure} 

\begin{figure}[htb] 
\begin{center} 
\includegraphics[width=\columnwidth]{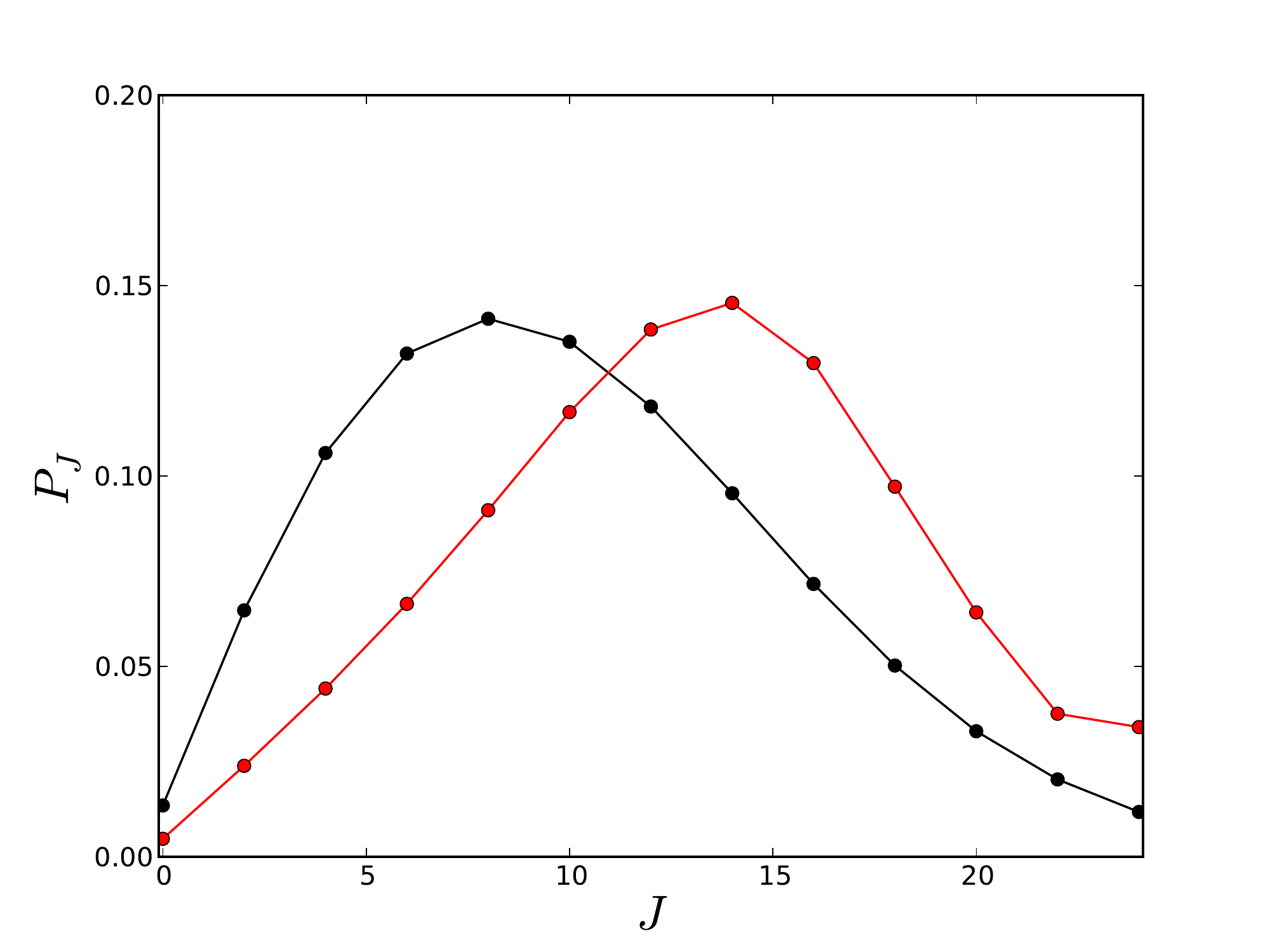}
\caption{ 
\label{Jdist} Angular momentum decomposition of the
wave function of the lighter fragment.  The black and red circles are probabilities
at $t= 0$ and $500$ fm/c, respectively.
\label{J-dist}
}
\end{center} 
\end{figure}

We next carry out the numerical 
integration of Eq. (\ref{Ht}) with the $V_Q(t)$ from Fig. (1).  Fig. 2
shows the expectation value of the quadrupole moment as a function of 
time.  Fig. 3 shows the corresponding average angular momentum.  One 
sees that it increases significantly in the first few hundred fm/c
reaching a steady value state by $\sim$500 fm/c.  
Fig. \ref{Jdist} compares the final $J$ distribution 
with the initial distribution. The position of the peak
is shifted upward by about 5 units, and the average by
3 units. 

{\it Parameter variation}  There is abundant evidence from the
gamma cascade in the excited daughter nuclei \cite{ca09,st14,ra14} that
the average initial angular momentum is much smaller that what
we found with our parameterization based 
on $\beta = 0.5$.  It would therefore be informative to examine
the dependence on the parameter values.  Table I shows the
effect of varying some of model assumptions.  The first entry
A corresponds to the assumptions underlying the example from
the previous section.  The first parameter we varied is
the moment of inertia, changing it by a factor of 2.  There is
no change increasing it, and a mild decrease when it is smaller.
Thus, the results are rather insensitive to $\cal I$, provided
it is not far from the physical value.  
\begin{table}[htb] 
\begin{center} 
\begin{tabular}{|c|ll|cl|c|} 
\hline 
&  &     &       \multicolumn{2}{c|}{$\langle J^2\rangle^{1/2}$} &\\ 
&$Q_e$  & ${\cal I}/\hbar^2$   & 0  & 500   & $\Delta J$\\
\hline 
A&470 e-b  & 17.5 Mev$^{-1}$   &  11.8    & 14.5  & 2.7\\
B&470  & 35.0   &  11.8    & 14.5 & 2.7\\
C&470  & 8.8   &  11.8    & 13.5 & 1.7\\
D&470  & 17.5   &  6.0  & 8.7   & 2.7 \\
E&235  & 16.1   &  6.0  & 7.4  &  1.4 \\
\hline 
\end{tabular} 
\caption{
\label{various}
Angular momentum at $t=0$ and $t=500$ fm/c  under various sets of 
Hamiltonian parameters and initial wave function.  A: the set
with results shown in Figs. (6-8):  B,C: same as A except for
moment of inertia;  D: same as A except for initial angular momentum
distribution;  E: Hamiltonian parameters given by Eq. (2,3,6) and (7) for
$\beta = 0.25$.
}
\end{center} 
\end{table} 
The next change we considered is in the initial angular momentum
distribution.  Our distribution was determined assuming that
the deformation axis of the daughter nucleus was perfectly
aligned with the fission axis.  In fact early theory based on
classical concepts sometimes invoked excitation of a bending
degree of freedom to account for the angular of the fragments.
In our formalism, inclusion of amplitudes of configurations
such as depicted in Fig. 3 of Ref. \cite{ra69}  would delocalize the alignment of
the axes and thereby lower the angular momentum content of
the deformed wave packet.  To explore this degree of freedom,
we arbitrarily decreased the average angular momentum by a 
factor of two, changing $\sigma$ in Eq. (\ref{PJ}) by about that amount.
The results are shown in entry D.  Both angular momenta are
much smaller, but the increase during the post-scission
acceleration remains the same.  The last variation we
examined is to reduce $\beta$ by a factor of two and change
all the Hamiltonian parameters accordingly.  The resulting
reorientation effect is now reduced, adding only 1.4 units
to the average.

{\it Conclusion} The main determinant of the angular momentum gain of fission
fragments during the post-scission acceleration 
is their initial deformation and orientational alignment.
The magnitude of the gain is of the order 1-3 units, which 
small not completely 
negligible compared to values 6-12 units at the scission point.
We conclude that the mechanism studied here would be significant if
the theory of the angular momentum generation were  reliable to the 10\%
level. However, in view of the much larger uncertainties in present theory,
the reorientation effect can probably be ignored.  The computer codes
used to generate Figs. 2-4 are provided in the Supplementary Material
\cite{supp}.

{\it Acknowledgment.} 
The author thanks T. Kawano for discussions that led to this work.

\end{document}